\documentclass[aps,prc,twocolumn,amsmath,amssymb,superscriaddress,showpacs,showkeys]{revtex4}
\usepackage{dcolumn}
\usepackage{graphicx,color}
\usepackage{textcomp}
\begin {document}
  \newcommand {\nc} {\newcommand}
  \nc {\beq} {\begin{eqnarray}}
  \nc {\eeq} {\nonumber \end{eqnarray}}
  \nc {\eeqn}[1] {\label {#1} \end{eqnarray}}
  \nc {\eol} {\nonumber \\}
  \nc {\eoln}[1] {\label {#1} \\}
  \nc {\ve} [1] {\mbox{\boldmath $#1$}}
  \nc {\ves} [1] {\mbox{\boldmath ${\scriptstyle #1}$}}
  \nc {\mrm} [1] {\mathrm{#1}}
  \nc {\half} {\mbox{$\frac{1}{2}$}}
  \nc {\thal} {\mbox{$\frac{3}{2}$}}
  \nc {\fial} {\mbox{$\frac{5}{2}$}}
  \nc {\la} {\mbox{$\langle$}}
  \nc {\ra} {\mbox{$\rangle$}}
  \nc {\etal} {\emph{et al.\ }}
  \nc {\eq} [1] {(\ref{#1})}
  \nc {\Eq} [1] {Eq.~(\ref{#1})}
  \nc {\Ref} [1] {Ref.~\cite{#1}}
  \nc {\Refc} [2] {Refs.~\cite[#1]{#2}}
  \nc {\Sec} [1] {Sec.~\ref{#1}}
  \nc {\chap} [1] {Chapter~\ref{#1}}
  \nc {\anx} [1] {Appendix~\ref{#1}}
  \nc {\tbl} [1] {Table~\ref{#1}}
  \nc {\Fig} [1] {Fig.~\ref{#1}}
  \nc {\ex} [1] {$^{#1}$}
  \nc {\Sch} {Schr\"odinger }
  \nc {\flim} [2] {\mathop{\longrightarrow}\limits_{{#1}\rightarrow{#2}}}
  \nc {\textdegr}{$^{\circ}$}
  \nc {\IR} [1]{\textcolor{red}{#1}}
  \nc {\IB} [1]{\textcolor{blue}{#1}}

\title{The ratio method: a new tool to study one-neutron halo nuclei}

\author{P.~Capel}
\email{pierre.capel@centraliens.net}
\affiliation{Physique Nucl\'eaire et Physique Quantique (C.P. 229), Universit\'e Libre de Bruxelles (ULB), B-1050 Brussels, Belgium}

\author{R.~C.~Johnson}
\email{r.johnson@surrey.ac.uk}
\affiliation{Department of Physics, University of Surrey, Guildford GU2 7XH, United Kingdom}
\affiliation{National Superconducting Cyclotron Laboratory and Department of Physics and Astronomy, Michigan State University, East Lansing, Michigan 48824, USA}

\author{F.~M.~Nunes}
\email{nunes@nscl.msu.edu}
\affiliation{National Superconducting Cyclotron Laboratory and Department of Physics and Astronomy,Michigan State University, East Lansing, Michigan 48824, USA}

\date{\today}

\begin{abstract}
Recently a new observable to study halo nuclei was introduced, based on the ratio between breakup  and elastic angular cross sections.
This new observable is shown by the analysis of specific reactions to be independent of the reaction mechanism and to provide
nuclear-structure information of the projectile. Here we explore the details of this ratio method, including
the sensitivity to binding energy and angular momentum of the projectile. We also study the reliability of the method
with breakup energy. Finally, we provide guidelines and specific examples for experimentalists who wish to apply this method.
\end{abstract}


\maketitle
%


\section{Introduction}
One of the most intriguing phenomenon revealed by the studies with rare isotope beams is that of halo nuclei \cite{jensen06}.
Since the early experiments on reaction cross sections \cite{tanihata85}, we have built a good understanding
of the exotic features that results from the proximity to threshold and the absence of repulsive barriers \cite{jensen00}.
For very loosely-bound nucleons, which do not suffer the constraint of the centrifugal or Coulomb barrier, the wavefunctions develop long tails, extending well into the classically forbidden region. A primary signature of the halo phenomenon
has been the sudden increase of the matter radius within a given isotopic chain \cite{c22}. In this respect, precision measurements of nuclear radii in traps open new possibilities \cite{he6radius,li11radius}. Narrow momentum distributions are also an indication of the large spatial extension of the wavefunctions \cite{zahar93,bazin98}. While identifying a halo can be  in itself a challenge, one would also like to have a better understanding of the structure of the valence orbital and its wavefunction, such as done in $(d,p)$ studies \cite{be10dp}. A new observable based on a ratio of cross sections, and therefore referred to as the ratio method \cite{ratio-plb}, offers this possibility. In this work, we explore the ratio method in detail and provide guidelines for its application.

Over the last two decades, many nuclear halos have been discovered.
Examples include the one neutron halo of $^{11}$Be \cite{fukuda04} and the one proton halo $^8$B \cite{smedberg99},
and both  have been the focus of many studies. In the recent years, experiments have been exploring heavier halos.  
After the measurement of $^{22}$C reaction cross section \cite{c22}, it was expected that $^{26}$O would also exhibit 
a two-neutron halo structure, but now it has become clear that this halo-to-be is indeed unbound \cite{o26}.
More recently, total reaction cross section measurements on $^{31}$Ne have been rather inconclusive
\cite{ne31-nakamura} as to whether $^{31}$Ne is a halo. A long sequence of theoretical studies \cite{ne31a,ne31b,ne31c,ne31d,ne31e} 
demonstrate that the structure information can depend strongly on the model used in the analysis of the reaction
and non integrated observables may be necessary. Another possibility of a halo has been identified in the Mg chain \cite{mg44} and others will surely surface with new technical developments.

As the mass increases, there are a limited number of isotopes for which $l=0,1$ orbitals are being filled in the ground state when the dripline is reached \cite{jensen00}. The halo phenomenon can however be more common in these systems because it also occurs in excited states. This is certainly known to be the situation for $^{11}$Be$(1/2_1^-)$ and  $^{17}$F$(1/2_1^+)$. Interest in determining halo properties of excited states continues \cite{ogloblin11} but no optimum probe has been found. In principle, the new observable here discussed, can be generalized to
characterize halo excited states too.

Halo nuclei are challenging from the nuclear structure point of view.
Fortunately, many-body methods are now able to adequately treat the asymptotic behavior of the
wavefunction of the loosely-bound nucleons \cite{radii-gfmc,radii-ncsm}.
Halo nuclei offer a unique testing ground for the understanding of the nuclear force and represent an opportunity to understand the density dependence of nuclear matter.

Due to its loosely-bound nature, the most likely method to explore halo structure is through breakup reactions. 
There have been important developments in the theory for the breakup of halo nuclei  (see Refs.~\cite{cdcc,cdcc2,cdcc3,esbensen-tdse,dea,dea2}), 
and today it is understood that non-perturbative non-classical approaches, which treat nuclear and Coulomb on equal footing, are needed to obtain reliable angular distributions \cite{hussein06,capel12}.
What is critical to understand is that the information to be extracted is model dependent, whether the process is inclusive or exclusive, whether it is nuclear or Coulomb dominated. When non-perturbative methods are used to solve the problem, the model dependence arises primarily from uncertainties in the core-target interaction \cite{capel04}. This interaction is often poorly known, specially when the core itself is radioactive, and plays a very important role in the breakup process.

In Ref.~\cite{capel10}, Capel \etal realized that the elastic and breakup angular distributions of halo nuclei exhibit very similar features. In Ref.~\cite{ratio-plb} the idea of taking the ratio of these angular distributions is introduced, drawing on the Recoil Excitation and Breakup model (REB) developed earlier \cite{johnson97, proc97, joh98}.
The main advantage of this new reaction observable is that it is nearly independent of the reaction mechanism and
that its sensitivity to the core-target interaction is strongly reduced.
The first application presented in Ref.~\cite{ratio-plb} is very encouraging.
Here we explore this ratio method in more detail. 

This paper is organized in the following manner.
In \Sec{theory} we provide the theoretical framework.
A discussion of various possible ratios is presented in \Sec{ratios}.
In \Sec{analysis} we demonstrate the validity of the ratio method and its range of validity.
In \Sec{structure} we study the structure information contained in this new observable.
Finally, conclusions are drawn in \Sec{conclusion}.


\section{Theoretical framework}\label{theory}

Describing the reaction of an exotic projectile $P$ impinging  on a target $T$ is a
complex many-body problem. While many-body techniques have made important advances to handle
a number of reactions, for example capture reactions on light nuclei \cite{navratil10,navratil11,navratil12} and nucleon elastic scattering \cite{hagen12}, these techniques are not able to handle the general heavy-ion reaction $P+T$  problem. In that
case, it is imperative to identify the relevant degrees of freedom and address the problem within
a few-body framework.

If the nucleus under study (the projectile) has a one-neutron halo, there is a strong decoupling of the
core degrees of freedom from the valence neutron \cite{jensen06}.
The projectile can then be described as a two-body system: a neutron $n$ loosely bound to a core $c$, i.e.  $P=c+n$.
In such a scenario, one can describe the states of the system with a mean field $V_{cn}$ that reproduces basic features, such as binding energy and radius, excited states or resonances, etc.
Assuming the target is well bound and focusing on elastic breakup only,
the implicit inclusion of target excitation through the imaginary part of optical potentials should be sufficient.
In this case, one can reduce the reaction $P+T$ to a three-body problem.
This is the approach considered here. To retain simplicity in our discussion, 
we consider the target and the core to be structureless and of spin zero although all the formalism can be extended to include target and core spins.

\subsection{The three-body model for nuclear reactions}\label{model}

In this model, the projectile $P=c+n$
is described by the Hamiltonian
\begin{equation}
H_0=-\frac{\hbar^2}{2\mu}\Delta+V_{cn}(\ve{r}),
\label{h-proj}
\end{equation}
where $\ve{r}$ is the $c$-$n$ relative coordinate and $\mu=m_nm_c/(m_n+m_c)$ is the $c$-$n$ reduced mass,
with $m_n$ and $m_c$ the masses of the neutron and of the core, respectively.
As mentioned above, $V_{cn}$ is a phenomenological mean field that captures
some essential aspects of the halo projectile and in  principle could be microscopically derived
(we assume it has central and spin-orbit terms, see Appendix~\ref{numerics} for details).

The eigenstates of $H_0$ are solutions of
\beq
H_0\phi_{ljm}(E,\ve{r})=E \phi_{ljm}(E,\ve{r}),
\eeqn{e2}
where $E$ is the $c$-$n$ relative energy. The total angular momentum $j$
results from the coupling of the orbital angular momentum $l$ and the spin of the valence neutron;
$m$ is its projection.
Negative-energy states correspond to $c$-$n$ bound states. They are normalized to unity.
We denote by $\{\phi_{l_ij_im_i}\}_{i=0,1,\ldots}$ these bound states of energy $E_i<0$,
with $i=0$ corresponding to the projectile ground state, $i=1$ to its first excited (bound) state, etc.
Positive-energy states describe the $c$-$n$ continuum. Their radial part is normalized according to
\beq
u_{lj}(E,r)\flim{r}{\infty}\sqrt{\frac{2\mu}{\pi\hbar^2 k}}\left [\cos\delta_{lj}F_l(kr)+\sin\delta_{lj}G_l(kr)\right],
\eeqn{e3}
where $k=\sqrt{2\mu E/\hbar^2}$ is the wave number, $\delta_{lj}$ is the nuclear phase shift at energy $E$, and 
$F_l$ and $G_l$ are regular and irregular Coulomb functions, respectively \cite{AS70},  taken for zero Sommerfeld parameter.
This normalization has been chosen so that
\beq
\langle\phi_{ljm}(E)|\phi_{ljm}(E')\rangle=\delta(E-E').
\eeqn{e4}

With this two-body description for the projectile, the $P$-$T$ collision reduces
to a three-body problem whose Hamiltonian reads
\begin{equation}
H_{\rm 3b}(\ve{R},\ve{r})= \hat T_{\ve{R}} + H_0(\ve{r}) + U_{cT}(\ve{R}_c) + U_{{\rm n}T}(\ve{R}_n),
\label{h3b}
\end{equation}
where $\ve R$ is the coordinate of the projectile center of mass relative to the target.
In \Eq{h3b}, additional optical potentials have been introduced to describe the scattering of the core
off the target $U_{cT}$ and the neutron off the target $U_{nT}$.
These optical potentials are typically phenomenological and contain an important imaginary term
to account for other reaction channels not explicitly included in this description.
Since they are not uniquely defined, these potentials may induce significant uncertainties
in the analysis of the reactions modeled within this framework \cite{capel04}.

In order to study the reactions of $P$ on $T$ we need to solve the three-body \Sch equation
\begin{equation}
H_{\rm 3b}\Psi(\ve{R},\ve{r})=E_{\rm tot}\Psi(\ve{R},\ve{r}) .
\label{3beq}
\end{equation}
As customary, we align the initial momentum $\ve K_0$ with the $Z$ axis and
assume the projectile to be initially in its ground state, so that:
\begin{equation}
\Psi(\ve{R},\ve{r})\flim{Z}{-\infty}e^{iK_0Z}\phi_{l_0j_0m_0}(\ve{r}).
\label{boundary}
\end{equation}
The total energy of the system is then given by $E_{\rm tot}=\hbar^2K_0^2/2\mu_{PT}+E_0$,
where $\mu_{PT}$ is the $P$-$T$ reduced mass.

\subsection{The Dynamical Eikonal Approximation}\label{dea}

It is important to identify a method for solving the three-body problem \eq{3beq},
that reliably describes elastic and breakup of loosely-bound nuclei.
While the momentum-space integral Faddeev method \cite{faddeev,faddeev2} is considered exact, its present implementation
is limited to $d+T$ reactions where the target charge is $Z_T \leq 20$. The Continuum Discretized Coupled Channel (CDCC) method
\cite{cdcc,cdcc2,cdcc3} compares well with the Faddeev method \cite{upadhyay}, however it is still computationally intensive.
A detailed comparison of the Dynamical Eikonal Approximation (DEA) \cite{dea,dea2} with CDCC shows that DEA is a very good approximation to the problem for
beam energies $E \geq 40$ MeV/nucleon \cite{capel12}.
Because this method is less computationally intensive, DEA is used in \Ref{ratio-plb} and here
to demonstrate the ratio method. 

A well known and useful approach to reactions at high energies is the eikonal approximation \cite{Glauber,dea,dea2}.
Motivated by the boundary form \eq{boundary}, the three-body wave function $\Psi(\ve{R},\ve{r})$ is factorized as
\beq
\Psi^{\rm DEA}(\ve{R},\ve{r})=
e^{i K_0 Z}\widehat\Psi(\ve{R},\ve{r}).
\eeqn{deawf}
At high energies, one expects a weak dependence on $\ve{R}$ of $\widehat\Psi$. 
Using the factorization \eq{deawf} in \Eq{3beq} and neglecting second-order derivatives of $\widehat\Psi$ with respect to $\ve R$,
we obtain \cite{dea2}
\beq
\lefteqn{i\frac{\hbar^2K_0}{\mu_{PT}}\frac{\partial}{\partial Z}
\widehat\Psi(Z,\ve{b},\ve{r})}\nonumber\\
&=& \left[(H_0-E_0)+U_{cT}(\ve{R}_c)+U_{nT}(\ve{R}_n)\right]
\widehat{\Psi}(Z,\ve{b},\ve{r}),
\eeqn{dea-eq}
where $Z$ and $\ve{b}$ are the longitudinal and transverse components of $\ve{R}$, respectively.
In the standard eikonal implementation, a subsequent adiabatic approximation is performed to solve \Eq{dea-eq}.
That approximation corresponds to neglect the excitation energy of the projectile compared to the beam energy.
In DEA, no such an adiabatic approximation is made and \Eq{dea-eq} is solved numerically for each $\ve{b}$
imposing the condition: $\widehat{\Psi}(Z\rightarrow-\infty,\ve{b},\ve{r})=\phi_{l_0j_0m_0}(\ve{r})$, in agreement with
condition \eq{boundary}.
The S-matrix is then extracted from the asymptotic behavior $\widehat{\Psi}(Z\rightarrow+\infty,\ve{b},\ve{r})$
as detailed in \Ref{dea2}.
Note that since this does not imply any semiclassical hypothesis, DEA is a fully quantal model \cite{dea2,capel12}.

\subsection{The recoil excitation and breakup model}

In the nineties, Johnson {\it et al.} realized that a simple factorization of the scattering amplitude
can be obtained when a neutron halo projectile interacts with the target \cite{johnson97}.
The key ingredients to the so-called Recoil Excitation and Breakup (REB) model are
i) neglecting the valence particle's interaction with the target, and 
ii) assuming the excitation energy of the projectile is small compared to the beam energy (the adiabatic approximation).
When these two conditions are satisfied, the elastic-scattering cross sections becomes \cite{johnson97, proc97}: 
\begin{equation}
\left( \frac{d \sigma}{d \Omega} \right)_{\rm el} = | F_{0,0}(\ve Q)|^2 \;  \left( \frac{d \sigma}{d \Omega} \right)_{\rm pt} 
\label{reb-xs}
\end{equation}
where $| F_{0,0}(\ve Q)|^2$ is a form factor accounting for the extension of the halo [see \Eq{reb-ff} below], and
$\left( \frac{d \sigma}{d \Omega} \right)_{\rm pt}$ is a cross section for a point-like projectile with mass $\mu_{PT}$,  scattered by the core-target interaction $U_{cT}$.
The relation \eq{reb-xs} is often mistaken for the first-order perturbation theory although it does not involve
the Born approximation.
Note that $\left( \frac{d \sigma}{d \Omega} \right)_{\rm pt}$ is similar to the experimental core-target elastic scattering, but for a different projectile mass.

The form factor is defined by:
\beq
|F_{0,0}(\ve Q)|^2=\frac{1}{2j_0+1}\sum_{m_0}\left|\int|\phi_{l_0j_0m_0}(\ve{r})|^2 e^{i\ve{Q\cdot r}}d\ve{r}\right|^2,
\eeqn{reb-ff}
and represents the Fourier transform of the halo ground state density.
Here $\ve Q = \frac{m_n}{m_c+m_n} (K_0\ve{\widehat{Z}} - \ve K')$ is proportional to the momentum transferred
during the scattering process. It modulates the diffraction pattern contained in the point-like cross section, determining which are the relevant scattering angles to be considered in the process
\beq
Q=2\frac{m_n}{m_c+m_n} K_0 \sin(\theta/2).
\eeqn{eQ}

In \Ref{capel10}, it was realized that the elastic and breakup cross sections have similar diffraction patterns,
a fact only fully understood with the subsequent work on the ratio method \cite{ratio-plb}.
In \Ref{ratio-plb}, we used the fact that the factorization \eq{reb-xs} can be generalized to angular distributions
for the excitation of the projectile to any of its state, either bound or not \cite{proc97,joh98}.
For inelastic scattering with excitation to bound state $i>0$, we can define the form factor
\beq
\lefteqn{|F_{i,0}(\ve Q)|^2}\nonumber\\
&=& \frac{1}{2j_0+1}\sum_{m_0}\sum_{m_i}\left| \int\phi_{l_ij_im_i}(\ve{r}) \phi_{l_0j_0m_0}(\ve{r})
e^{i\ve{Q\cdot r}}d\ve{r}\right|^2,
\eeqn{e11}
while for breakup to energy $E$, we use the form factor
\beq
\lefteqn{|F_{E,0}(\ve Q)|^2}\nonumber\\
&=& \frac{1}{2j_0+1}\sum_{m_0}\sum_{ljm} \left| \int\phi_{ljm}(E,\ve{r}) \phi_{l_0j_0m_0}(\ve{r})
e^{i\ve{Q\cdot r}}d\ve{r}\right|^2,
\eeqn{reb-ff2}
where $\phi_{ljm}(E,\ve{r})$ is the eigenstate of $H_0$ at positive energy $E$ in the partial wave $ljm$ [see \Eq{e2}].
The REB prediction for the inelastic cross section, i.e. the angular distribution for the projectile excited to state $i$ while scattered in direction $\Omega$ reads
\beq
\left( \frac{d \sigma_i}{d \Omega} \right)_{\rm inel} = | F_{i,0}(\ve Q)|^2 \;  \left( \frac{d \sigma}{d \Omega} \right)_{\rm pt}.
\eeqn{reb-inel}
Similarly, we get the following angular distribution for breakup, i.e.\ the cross section for the projectile being broken up at an energy $E$ in the $c$-$n$ continuum with its center of mass scattered in direction $\Omega$
\begin{equation}
\left( \frac{d \sigma}{dE d \Omega} \right)_{\rm bu} = |F_{E,0}(\ve Q)|^2 \; \left(\frac{d \sigma}{d \Omega} \right)_{\rm pt}.
\label{reb-br}
\end{equation}

Neglecting the small difference in magnitude between the outgoing momenta for elastic and inelastic processes, the point-like cross section $\left(d \sigma/d \Omega \right)_{\rm pt}$ is identical for all three processes \eq{reb-xs}, \eq{reb-inel}, and \eq{reb-br}. This first explains the result obtained in \Ref{capel10}, where it was observed that the angular distributions for elastic scattering and breakup exhibit very similar patterns. Indeed, most of the angular dependence of these cross sections is due to that point-like cross section. Second, the similarity of the expressions \eq{reb-xs}, \eq{reb-inel}, and \eq{reb-br} is at the core of the ratio method. If we now consider the ratio between Eqs.~\eq{reb-br} and \eq{reb-xs}, the point-like cross sections cancel out, leaving an observable which, within the REB model, is just the ratio of form factors
\beq
{\cal R}_{\rm el}(E,\ve Q) &=& \frac{(d\sigma/dEd\Omega)_{\rm bu}}{(d\sigma/d\Omega)_{\rm el}}\label{ratio-el}\\
 &\stackrel{\rm (REB)}{=}& \frac{|F_{E,0}(\ve Q)|^2}{| F_{0,0}(\ve Q)|^2}.
\eeqn{reb-el}
Therefore, according to the REB predictions, this ratio should be sensitive only to the structure of the projectile and be independent of the reaction mechanism. In particular, considering the ratio \eq{ratio-el} automatically removes the dependence on the core-target interaction, which is the most ambiguous input in reaction modeling.

\section{Ratios of cross sections}\label{ratios}

Before analyzing the structure content of this new observable, we should point out that identical cancellations of the point-like cross section can be obtained for the ratio of any linear combination of breakup, elastic- and inelastic-scattering angular distributions.
Therefore we consider here, in addition to ${\cal R}_{\rm el}$ \eq{ratio-el}, other options. Because in some halo systems, there is a nearby excited state, hard to disentangle from the ground state, the elastic and inelastic contributions may be easier to measure together. We then introduce the quasi-elastic ratio
\beq
{\cal R}_{\rm quasi}(E,\ve Q) &=& 
\frac{(d\sigma/dEd\Omega)_{\rm bu}}{(d\sigma/d\Omega)_{\rm quasi}}\label{ratio-quasi}\\
 &\stackrel{\rm (REB)}{=}&\frac{|F_{E,0}(\ve Q)|^2}{|F_{0,0}(\ve Q)|^ 2+\sum_{i>0}|F_{i,0}(\ve Q)|^ 2},
\eeqn{reb-quasi}
where $(d\sigma/d\Omega)_{\rm quasi}=(d\sigma/d\Omega)_{\rm el}+\sum_{i>0}(d\sigma_i/d\Omega)_{\rm inel}$.
Because for low $Q$, elastic scattering is dominant, adding the breakup does not make much difference to the ratio observables but simplifies the form factor dependence. Thus, we also consider
\beq
{\cal R}_{\rm sum}(E,\ve Q) &=& \frac{(d\sigma/dEd\Omega)_{\rm bu}}{(d\sigma/d\Omega)_{\rm sum}}\label{ratio-sum}\\
 &\stackrel{\rm (REB)}{=}&|F_{E,0}(\ve Q)|^2,
\eeqn{reb-sum}
where the summed cross section reads
\beq
\lefteqn{\left(\frac{d\sigma}{d\Omega}\right)_{\rm sum}}\nonumber\\
 &=&\left(\frac{d\sigma}{d\Omega}\right)_{\rm el}
+\sum_{i>0}\left(\frac{d\sigma_i}{d\Omega}\right)_{\rm inel}+\int \left(\frac{d\sigma}{dEd\Omega}\right)_{\rm bu} dE.
\eeqn{xs-sum} 
\begin{figure}[t]
\center
\includegraphics[width=8.7cm]{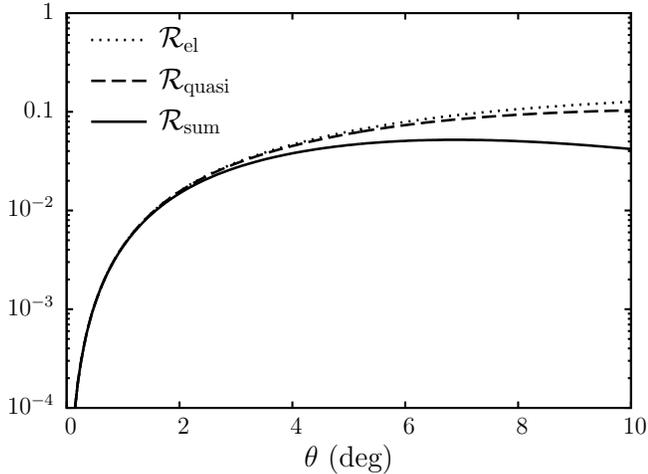}
\caption{Ratios \eq{reb-el}, \eq{reb-quasi} and \eq{reb-sum} suggested by the similarity between angular
distributions for elastic scattering and breakup. The calculations are performed within REB for \ex{11}Be impinging on Pb at 69~MeV/nucleon considering a \ex{10}Be-$n$ continuum energy $E=0.1$~MeV. }\label{ratios1}
\end{figure}
We compare in \Fig{ratios1} the REB prediction for ${\cal R}_{\rm el}$ \eq{reb-el}, ${\cal R}_{\rm quasi}$ \eq{reb-quasi} and ${\cal R}_{\rm sum}$ \eq{reb-sum} for the reaction of $^{11}$Be on $^{208}$Pb at 69 MeV/nucleon.
The transferred momentum $Q$ has been converted into the center-of-mass scattering angle following \Eq{eQ}.
As expected there is very little difference between ${\cal R}_{\rm el}$ and  ${\cal R}_{\rm quasi}$.
Adding the breakup angular distribution to the denominator modifies only the large-angle behavior of the ratio.
Other possibilities for the ratio are discussed in Appendix~\ref{other}.

After close analysis and a number of exploratory calculations, we found it optimal to consider the ratio ${\cal R}_{\rm sum}$. This ratio leads to the simplest REB prediction \eq{reb-sum} and is probably the easiest to measure experimentally.
In \Ref{ratio-plb} we quantified this ratio with DEA calculations that do not make the approximations of the REB model, and presented the argument that the REB approximations can be justified in realistic cases. Here we focus the discussion on the source of the small discrepancies found between DEA calculations and REB predictions and the structure information that can be extracted from ${\cal R}_{\rm sum}$.

\section{Analysis of the cross section ratio}\label{analysis}

\begin{figure}[ht]
\center
\includegraphics[width=8.7cm]{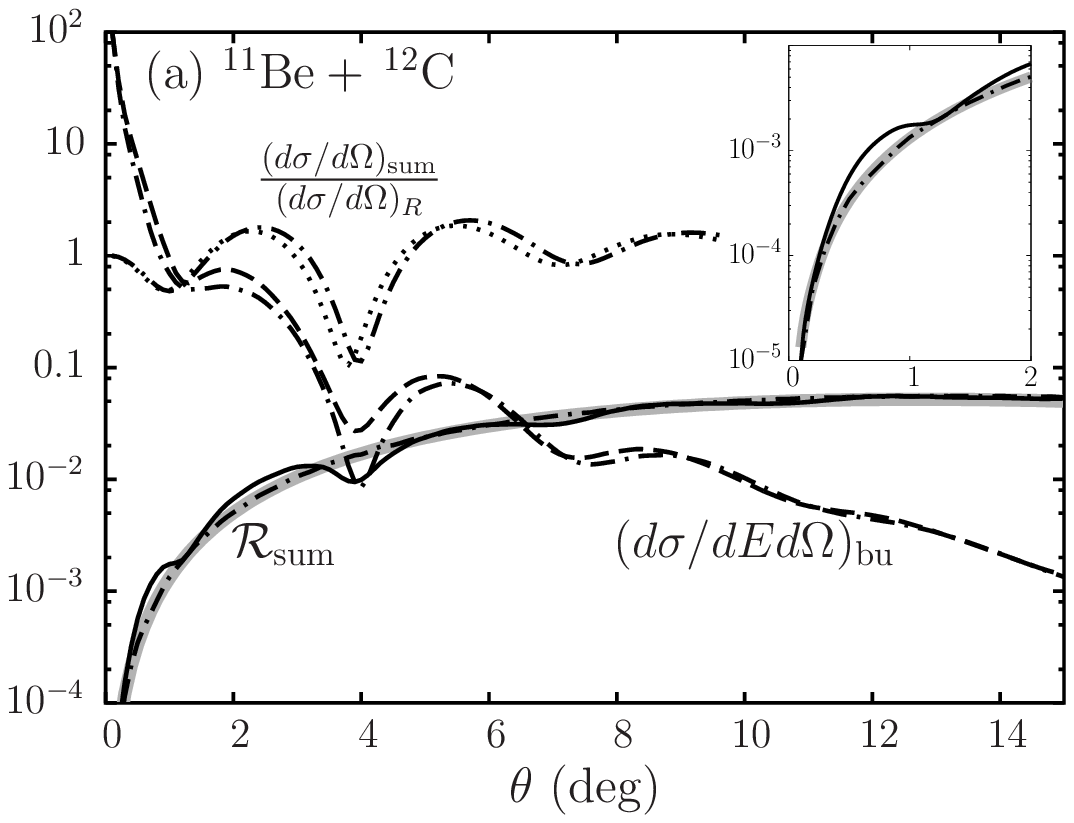}\\
\includegraphics[width=8.7cm]{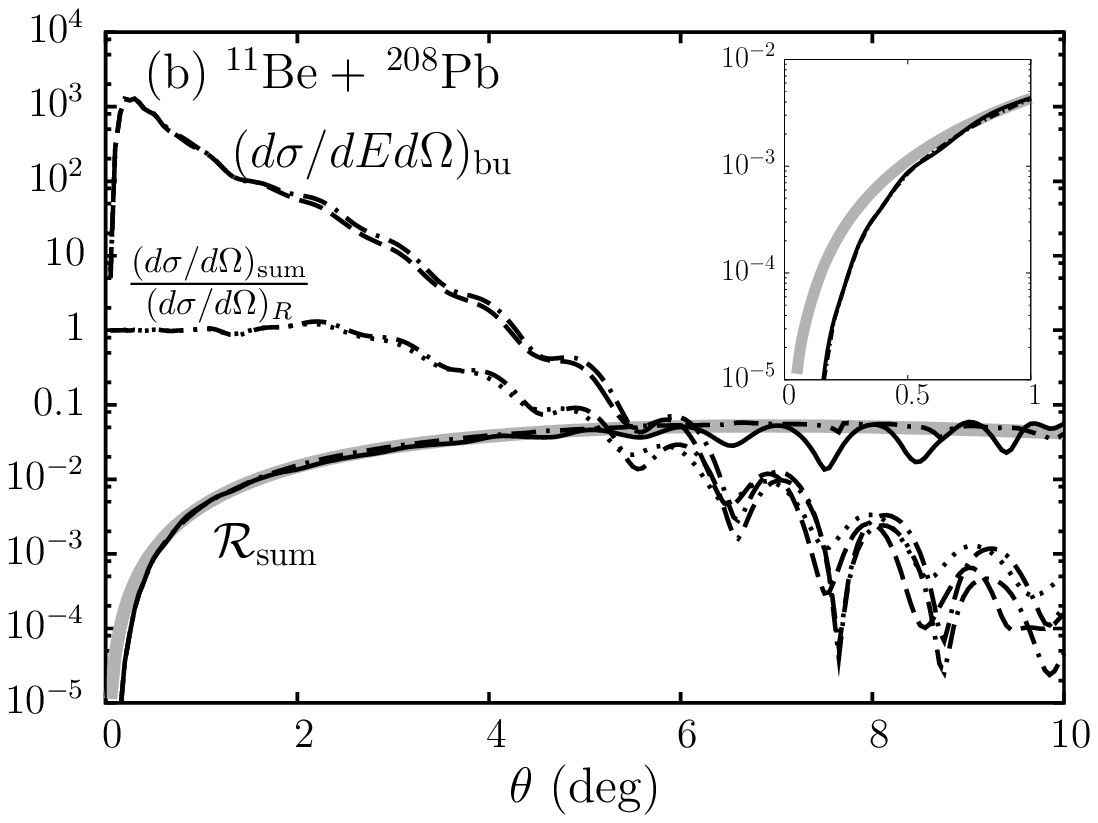}
\caption{Illustration of the ratio method for \ex{11}Be impinging on: (a) C at 67~MeV/nucleon, and (b) Pb at 69~MeV/nucleon.
Summed cross sections (dotted lines) and breakup angular distributions (dashed lines) computed within DEA are compared to their ratio ${\cal R}_{\rm sum}$ (thin solid lines), which is found in excellent agreement with its REB prediction $|F_{E,0}|^2$ (thick grey line).
Calculations with $U_{nT}=0$ are shown as dash-dotted lines.
The insets focus on the forward-angle behavior of the ratio.}\label{fig-vnt}
\end{figure}

For the purpose of illustration, we base our calculations on a concrete  reaction measured at RIKEN \cite{fukuda04}, namely the breakup of $^{11}$Be on C and Pb at 67 and 69~MeV/nucleon, respectively.
In our two-body description of the projectile, \ex{11}Be is seen as an inert \ex{10}Be core in its $0^+$ ground state, to which a neutron is bound by $E_0=-0.5$~MeV in the $1s_{1/2}$ orbit.
Unless mentioned otherwise, we take the same inputs as in \Ref{ratio-plb} (all interactions are provided in our Appendix~\ref{numerics})
and perform calculations within DEA, which is found in excellent agreement with these experimental data \cite{dea2}.

In Fig.~\ref{fig-vnt} we show the corresponding summed cross sections \eq{xs-sum} as a ratio to Rutherford (dotted lines),
the angular distributions for breakup at a continuum energy $E=0.1~{\rm MeV}$ in units b/MeV (dashed lines),
as well as their ratios ${\cal R}_{\rm sum}$ \eq{ratio-sum} in units MeV\ex{-1} (solid lines).
The continuum energy $E=0.1$ MeV is at this point arbitrary. We will discuss it in detail in \Sec{continuum}.
If all works well, that ratio ${\cal R}_{\rm sum}$ should agree with $|F_{E,0}|^2$ \Eq{reb-ff2}, as predicted by the REB model \eq{reb-sum} (thick grey line).
And, indeed, we find the agreement to be very good. In both cases, most of the angular dependence of the cross sections has been removed by taking their ratio, leaving a curve varying smoothly with the scattering angle $\theta$.
Moreover, this ratio lies nearly on top of its REB prediction.
As already pointed out in \Ref{ratio-plb}, this implies that the ratio ${\cal R}_{\rm sum}$ removes most of the dependence on the reaction mechanism and hence contains mostly structure information.
We note the presence of residual oscillations at forward angles for the C target and at larger angles for the Pb target. Note also the slower rise at the most forward angles in the latter case
(see the insets, which focus on the forward-angle region).
In the next subsections we explore the source for these small discrepancies.

\subsection{The REB cross sections}\label{reg}
\begin{figure}[t]
\center
\includegraphics[width=8.7cm]{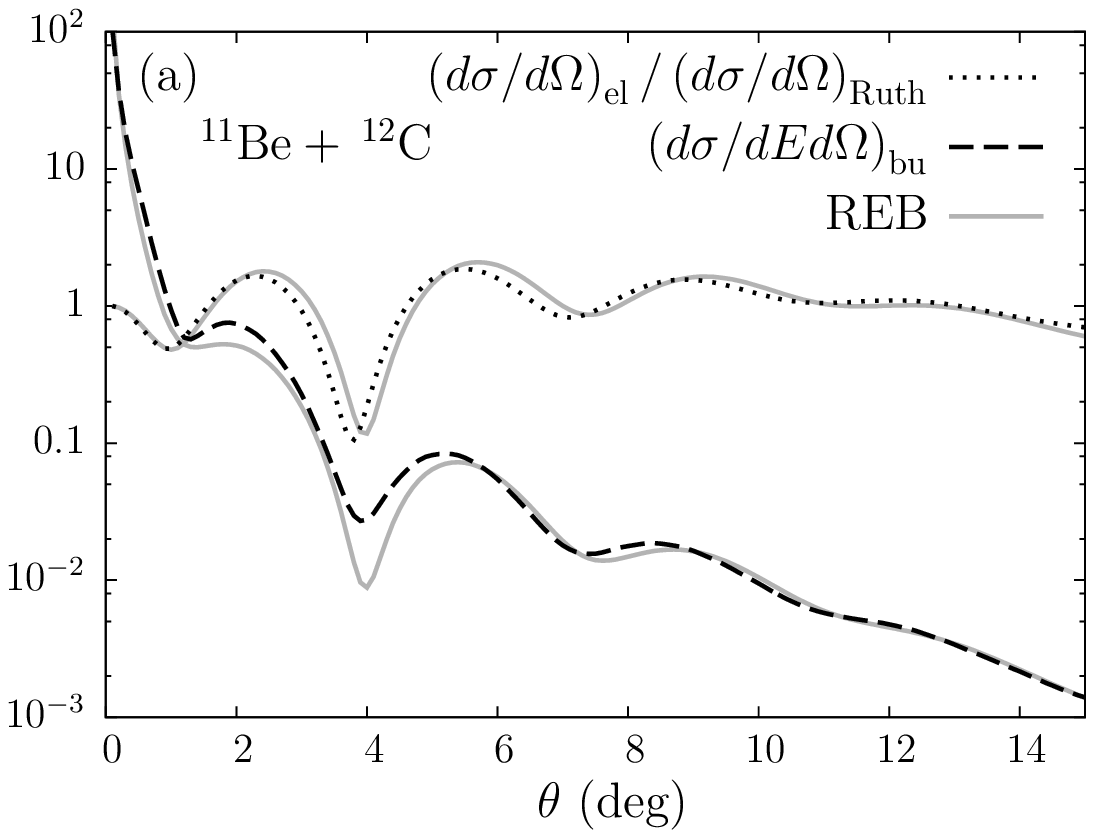}
\includegraphics[width=8.7cm]{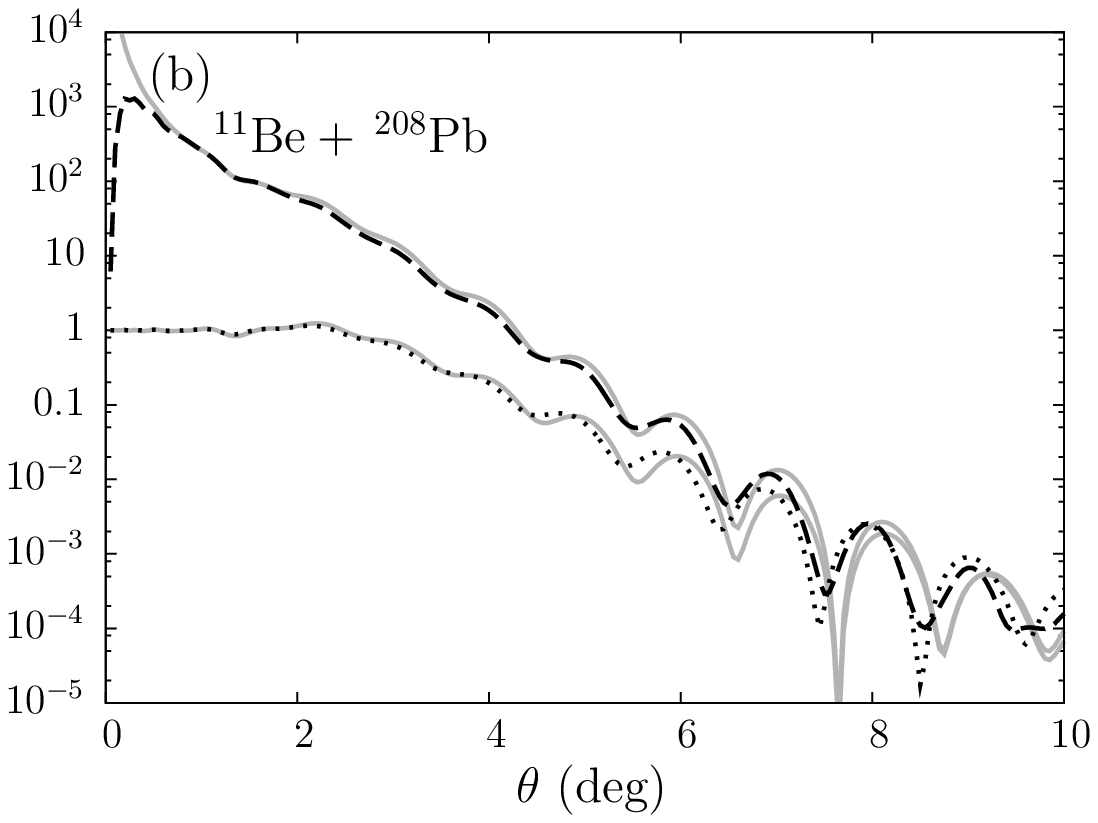}
\caption{Comparison of REB  and DEA predictions for the elastic  and breakup  angular distributions for $^{11}$Be: (a)  on $^{12}$C at 67 MeV/nucleon  and (b) on $^{208}$Pb at 69 MeV/nucleon. The dotted(dashed) line corresponds to the elastic(breakup) scattering within DEA and the grey lines to results with REB. }\label{fig-reb}
\label{reb-test}
\end{figure}
The validity of  the ratio (\ref{ratio-el})  depends crucially on the  equality of the two point-like cross sections in Eqs.~(\ref{reb-xs}) and (\ref{reb-br}). So here we test explicitly the validity of these equations.
In \Fig{reb-test} we compare the results of Eqs.~(\ref{reb-xs}) and (\ref{reb-br}) with those obtained in the full dynamical calculation (DEA), for our two examples, namely $^{11}$Be on $^{12}$C (a) and $^{11}$Be on $^{208}$Pb (b). The dotted and dashed lines correspond to the DEA angular distributions for the elastic and breakup cross sections, respectively. The grey lines are obtained with the factorization in Eqs.~(\ref{reb-xs}) and (\ref{reb-br}). For the light target, the REB follows closely the DEA result in both elastic and breakup processes but for a slight shift in the oscillatory pattern. The same can be seen for the Pb target with the exception of smaller angles. In this regime, the breakup cross section is not well described by REB. In the next two subsections we will discuss the two approximations present in the REB model and their imprint on the discrepancies seen in \Fig{reb-test}.

\subsection{Role of $U_{nT}$}\label{vnt}
The REB model neglects the contribution of $U_{nT}$ and this explains why the form factor $|F_{E,0}|^2$ is perfectly smooth, whereas the DEA ratio exhibits residual oscillations.
The neutron interaction with the target gives the projectile a minor kick that causes a slight shift in the
diffractive pattern, as already noted by Johnson \etal \cite{johnson97} and confirmed in \Fig{fig-reb}.
A careful analysis of the angular distributions shows that this shift depends slightly on the excitation energy of the projectile.  The oscillatory pattern in the dynamical calculations therefore differs between elastic, inelastic, and breakup cross sections, leading to the residual oscillations in their ratio. To confirm this analysis, we repeat the DEA calculations setting $U_{nT}=0$ (dash-dotted lines in Fig.~\ref{fig-vnt}).
The angular distributions obtained in this manner are exactly in phase and their ratios exhibit no residual oscillations.
These ratios are in perfect agreement with the REB predictions but for the very forward-angle region on the Pb target (see inset of \Fig{fig-vnt}(b)).
In that region, setting $U_{nT}=0$ does not improve the agreement between DEA and REB. The reason for that difference has to be looked for in the second ingredient of the REB model, i.\,e.\ the adiabatic approximation (see \Sec{adiabatic}).

\begin{figure}[t]
\center
\includegraphics[width=8.7cm]{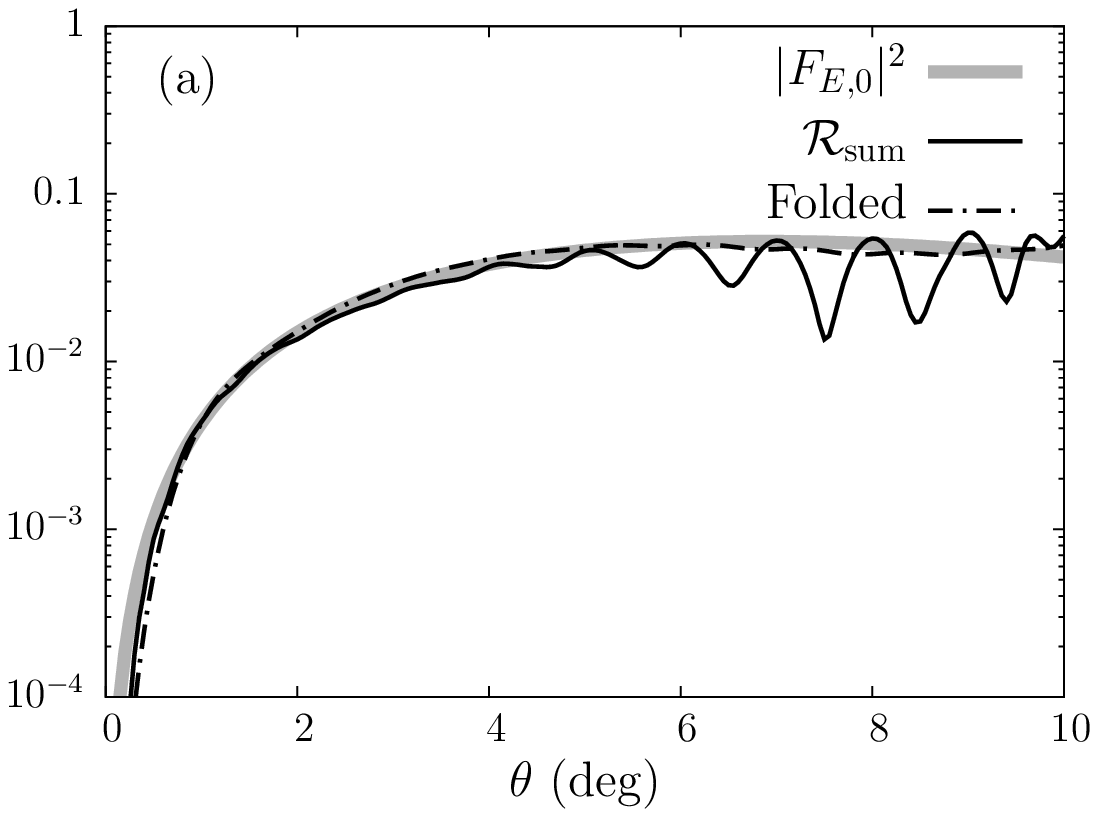}
\includegraphics[width=8.7cm]{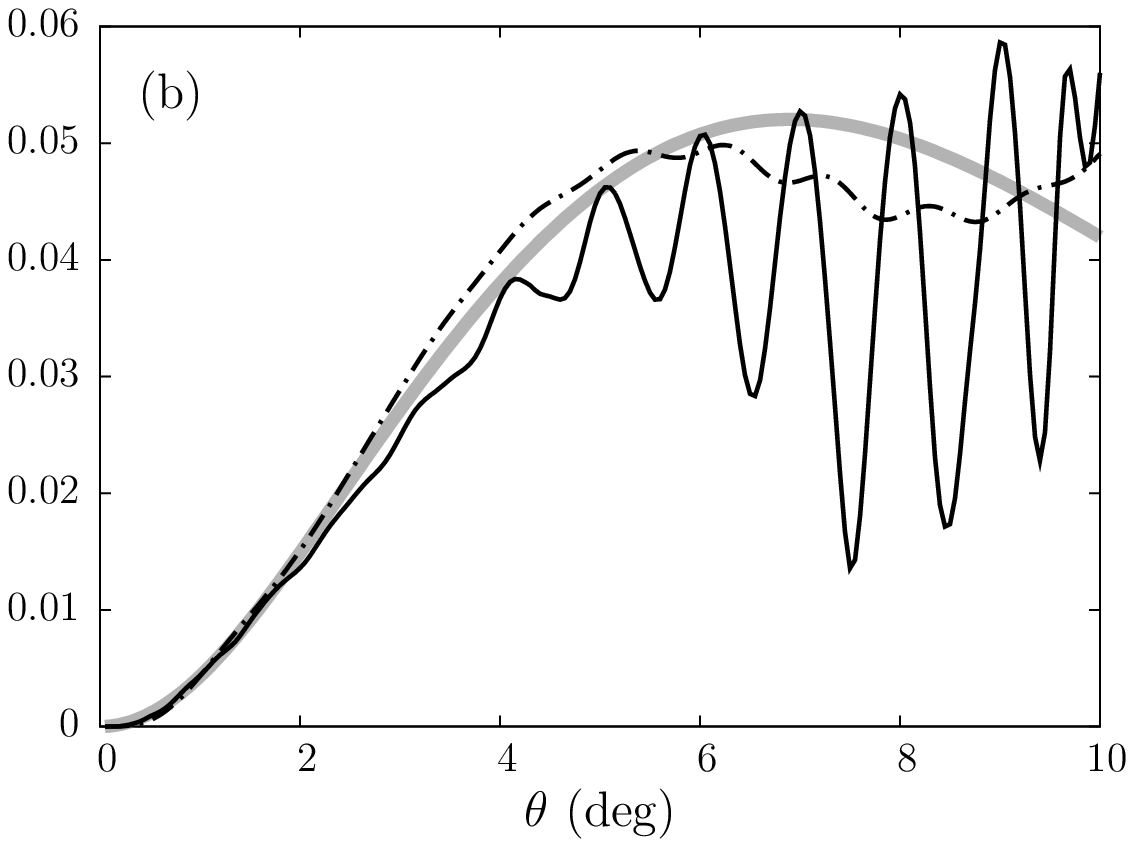}
\caption{Smoothing of the angular distributions by the folding with
experimental angular resolution (\ex{11}Be on Pb at 69~MeV/nucleon): (a) log plot; (b) linear plot.}\label{fig-resol}
\end{figure}

Even though $U_{nT}$ has an effect on the dynamics, when applying the ratio method to data it is likely that the residual oscillations will not be noticeable experimentally with current angular resolutions.
In \Fig{fig-resol} we show the ratio obtained with DEA (solid line), that predicted by the REB model (thick grey line) and that obtained after folding the DEA angular distributions with a typical experimental resolution (dash-dotted line). To this end, we convolute the theoretical cross sections with a Gaussian of standard deviation $0.41^\circ$, which corresponds to the angular resolution of the RIKEN experiment \cite{fukuda04}.
In \Fig{fig-resol}(a) we show the log plot, and to emphasize the difference we include \Fig{fig-resol}(b) with the corresponding linear plot.
As expected the convolution reduces the residual oscillations to the point where they would no longer be detectable.

\subsection{Role of the adiabatic approximation}\label{adiabatic}

In addition to neglecting $U_{nT}$, the REB model neglects the excitation energy of the projectile (adiabatic approximation).
This second approximation is responsible for the different slope of the ratio ${\cal R}_{\rm sum}$ and its REB prediction at the most forward angles on the Pb target (see inset of \Fig{fig-vnt}(b)).

The \Fig{reb-test}(b) shows very clearly that at very forward angles, the elastic scattering is well described by REB but the breakup cross section is not, introducing an unphysical divergence. One could arrive at these same conclusions directly by analysing the $Q$-dependence of Eqs.~(\ref{reb-xs}) and (\ref{reb-br}).
It is for this reason that the REB ratio \eq{reb-sum} is higher than the correct ${\cal R}_{\rm sum}$ at forward angles, as shown in the inset of \Fig{fig-vnt}(b).
The adiabatic---or sudden---approximation assumes a very brief interaction time with the target.
When the reaction is entirely dominated by the Coulomb interaction, which is the case for breakup on Pb at forward angles, it cannot be treated satisfactorily within the adiabatic approximation due to the infinite range of the Coulomb potential.
Note that the overestimation of the ratio by the REB is not observed for the carbon target. In that case the reaction is dominated by  short-ranged nuclear interactions, which allow us to rely on the adiabatic approximation \cite{dea2}.

This analysis shows that the effects of the adiabatic approximation upon the ratio are small and limited to the very forward angles for Coulomb-dominated reactions. Since cross sections in this region can hardly be measured, it is very unlikely that these effects will ever be noticeable. Nevertheless this analysis will help us understand differences between DEA calculations and REB predictions observed in later subsections.

\subsection{Independence of the ratio on the reaction process}\label{indep}
In \Ref{ratio-plb} we showed that the ratio obtained when considering the C target is identical to that obtained with a Pb target.
In other words, the new observable is independent of the reaction mechanism.
To appreciate this fact we emphasize the difference between the breakup and summed distributions obtained on C and on Pb (compare dotted and dashed lines in \Fig{fig-vnt}(a) and (b)).
Even though the cross sections are orders of magnitude apart, and their diffraction pattern is completely different, still the resulting ratio is very close to the form factor \Eq{reb-ff2} as predicted by REB.

\begin{figure}[t]
\center
\includegraphics[width=8.7cm]{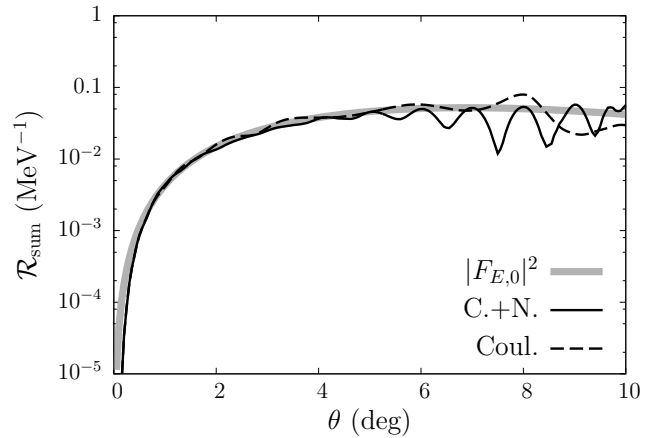}
\caption{Ratio  computed for \ex{11}Be on Pb at 69~MeV/nucleon using different interactions:
Coulomb plus nuclear (solid line) and purely Coulomb (dashed line) $P$-$T$ potentials.
}\label{fig-coul}
\end{figure}

In \Fig{fig-coul} we focus on the reaction on Pb and explore the interplay between Coulomb and nuclear interactions.
In addition to DEA calculations including Coulomb and nuclear interactions (C.+N., solid line),
we show results obtained from DEA calculations where only the Coulomb term of the $U_{cT}$ optical potential is considered (Coul., dashed line).
As already observed in \Ref{capel10}, the angular distributions vary strongly with the $P$-$T$ potential, indicating the sensitivity of the reaction mechanism to that potential choice.
Nevertheless, both ratios fall on top of the form factor, confirming the independence of the ratio to the reaction process.
The residual oscillations are significantly reduced when only the Coulomb interaction is present.
This is due to a much smoother behavior of the angular distributions when no nuclear interaction is included \cite{capel10}.

\begin{figure}[t]
\center
\includegraphics[width=8.7cm]{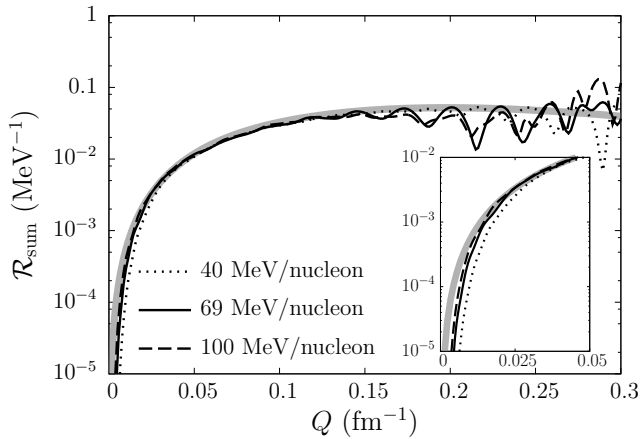}
\caption{Ratio computed on Pb at different beam energies.
}\label{fig-ebeam}
\end{figure}

To complete this analysis of the sensitivity of the ratio to the reaction mechanism, we now turn to its variation with the beam energy.
In \Fig{fig-ebeam}, $\cal{R}_{\rm sum}$ is plotted for \ex{11}Be impinging on Pb at 40, 69, and 100~MeV/nucleon.
As detailed in Appendix~\ref{numerics}, the optical potentials $U_{cT}$ and $U_{nT}$ are adapted to the beam energy, while the projectile description is kept unchanged.
To compare all three ratios to one another, they are plotted as a function of $Q$ \Eq{eQ}.
The most significant difference between all three calculations are observed at large $Q$, where the ratios exhibit residual oscillations.
As explained in \Sec{vnt}, they are due to $U_{nT}$, which varies with the beam energy.
Another, though smaller, difference is observed at very small $Q$, corresponding to very forward angle (see inset of \Fig{fig-ebeam}).
In that region, the DEA underestimates its REB prediction because of the adiabatic approximation (see \Sec{adiabatic}).
Since the REB approximation is more reliable at high beam energy, the agreement between DEA and REB improves at forward angles when larger energies are considered.

\subsection{Applicability to other one-neutron halo systems}\label{c19}
\begin{figure}[t]
\center
\includegraphics[width=8.7cm]{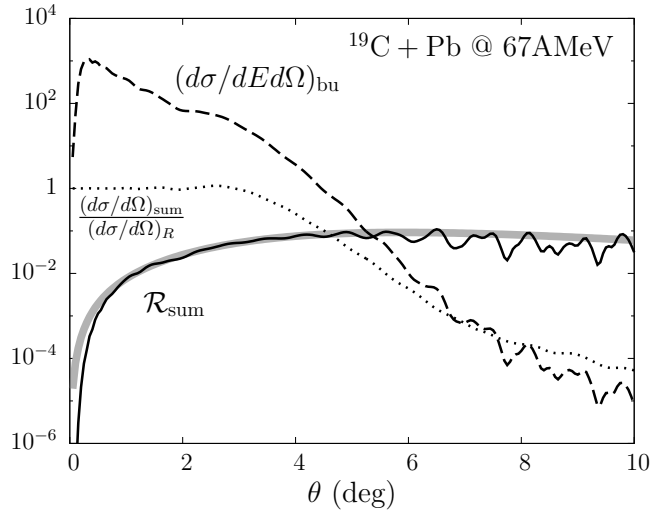}
\caption{Analysis of the ratio for \ex{19}C on Pb at 67~MeV/nucleon: summed cross section (dotted), breakup cross section (dashed) and ratio (solid) versus the REB prediction (thick grey).
}\label{fig-c19}
\end{figure}

To check the applicability of the ratio method to other halo nuclei, we study the case of \ex{19}C.
This one-neutron halo nucleus has been studied experimentally by various groups.
We choose here the conditions of the RIKEN experiment, i.\;e. performed at 67~MeV/nucleon on a lead target \cite{nakamura99}.
In \Fig{fig-c19}, the DEA summed (dotted line) and breakup (dashed line) cross sections are plotted as a function of the scattering angle $\theta$ of the \ex{18}C-$n$ center of mass together with the corresponding ratio ${\cal R}_{\rm sum}$ (solid line) and its REB prediction (thick gray line). Here we assumed the final breakup state to be a non-resonant state at $E=0.3$ MeV.
The results in \Fig{fig-c19} are very similar to those observed in \Fig{fig-vnt}(b) for \ex{11}Be: both angular distributions exhibit similar features that are mostly removed when taking their ratio.
This confirms the validity of the ratio method for other one-neutron halo projectiles.

\section{Structure information contained in the cross section ratio}\label{structure}

Now that we have a good understanding of the small discrepancies of the true ratio and the prediction from the REB model,
we can explore the structure information contained in this observable.
Having shown the ratio to be independent of the reaction process, we expect it to be more sensitive to the projectile structure than usual reaction observables.
Below we discuss the dependence on the binding energy of the halo neutron $E_0$, its orbital angular momentum $l_0$, the details of the $c$-$n$ radial wavefunction, and the final scattering state.
For this analysis, we stick to the collision of $^{11}$Be on Pb at 69/nucleon.
 
\subsection{Binding energy}
\begin{figure}[t]
\center
\includegraphics[width=8.7cm]{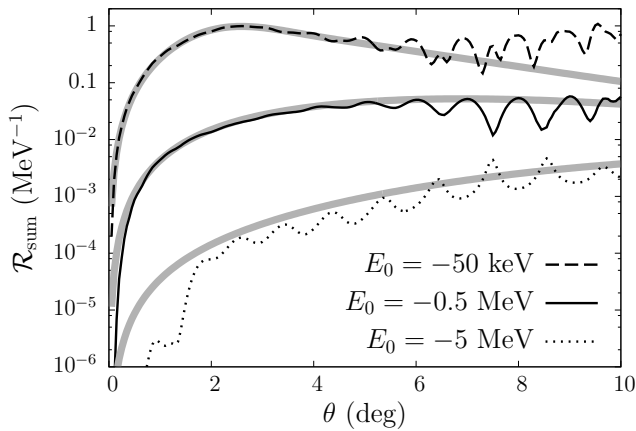}
\caption{Sensitivity of ratio ${\cal R}_{\rm sum}$ to the binding energy
of the projectile: $E_0=-50$~keV, $-0.5$~MeV, and $-5$~MeV.
}\label{fig-be}
\end{figure}

The ratio ${\cal R}_{\rm sum}$ is very sensitive to the one-neutron separation energy $E_0$.
Because the breakup cross section is larger for loosely bound systems, the magnitude of the ratio increases with decreasing binding \cite{ratio-plb}.
In \Fig{fig-be}, we show the ratio obtained for a $^{11}$Be-like system bound by 50~keV, 0.5~MeV, and 5.0~MeV, respectively.
They result from DEA calculations where the depth of the $^{10}$Be-$n$ interaction in the $s$ wave is adjusted to reproduce the appropriate one-neutron separation energy (see Appendix~\ref{numerics}).
Our results show that changing the binding energy by one order of magnitude produces a change in the ratio by two orders of magnitude.
Moreover, the shape of the ratio differs significantly from one binding energy to the other.
Looking into the details of the angular distributions one sees also that the agreement with the REB prediction deteriorates with increasing binding energy. This is to be expected since for large binding energy, the excitation energy needed for breakup is large and the adiabatic approximation is no longer justified. Nevertheless, it is clear that the cross section ratio provides a very accurate indirect measurement of the binding energy of the system.

\subsection{Orbital angular momentum}
\begin{figure}[t]
\center
\includegraphics[width=8.7cm]{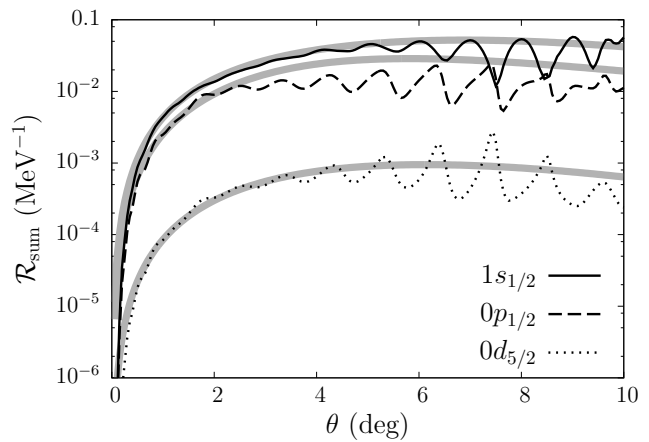}
\caption{Sensitivity of ratio ${\cal R}_{\rm sum}$ to the orbital angular momentum of
the halo neutron.
Calculations with a valence neutron bound to a \ex{10}Be core by $E_0=-0.5$~MeV in a $1s_{1/2}$, $0p_{1/2}$ or $0d_{5/2}$ orbital are compared to one another.
}\label{fig-ang}
\end{figure}

Next we investigate the dependence on the orbital angular momentum of the initial bound state $l_0$.
The $^{10}$Be-$n$ interaction is adjusted to reproduce a $^{11}$Be ground state at $E_0=-0.5$~MeV with, instead of the $1s_{1/2}$ configuration, a $0p_{1/2}$ and a $0d_{5/2}$ configuration (see Appendix~\ref{numerics}).
DEA calculations are repeated with these new interactions, and the resulting ratios are plotted in \Fig{fig-ang}.
Again we find that the cross section ratio is very sensitive to this property of the projectile initial state.
The magnitude of the ratio decreases with increasing angular momentum.
It is important to note that even though the magnitude for a 5~MeV bound $1s_{1/2}$ state (\Fig{fig-be}) is similar to that of a $0.5$ MeV $0d_{5/2}$ state (\Fig{fig-ang}), the shape of the distribution is very different, particularly the slope at larger angles.
This feature would make it possible to determine unequivocally both the binding energy and the angular momentum of the nucleus under inspection.

\subsection{Radial wave function}
\begin{figure}[ht]
\center
\includegraphics[width=8.7cm]{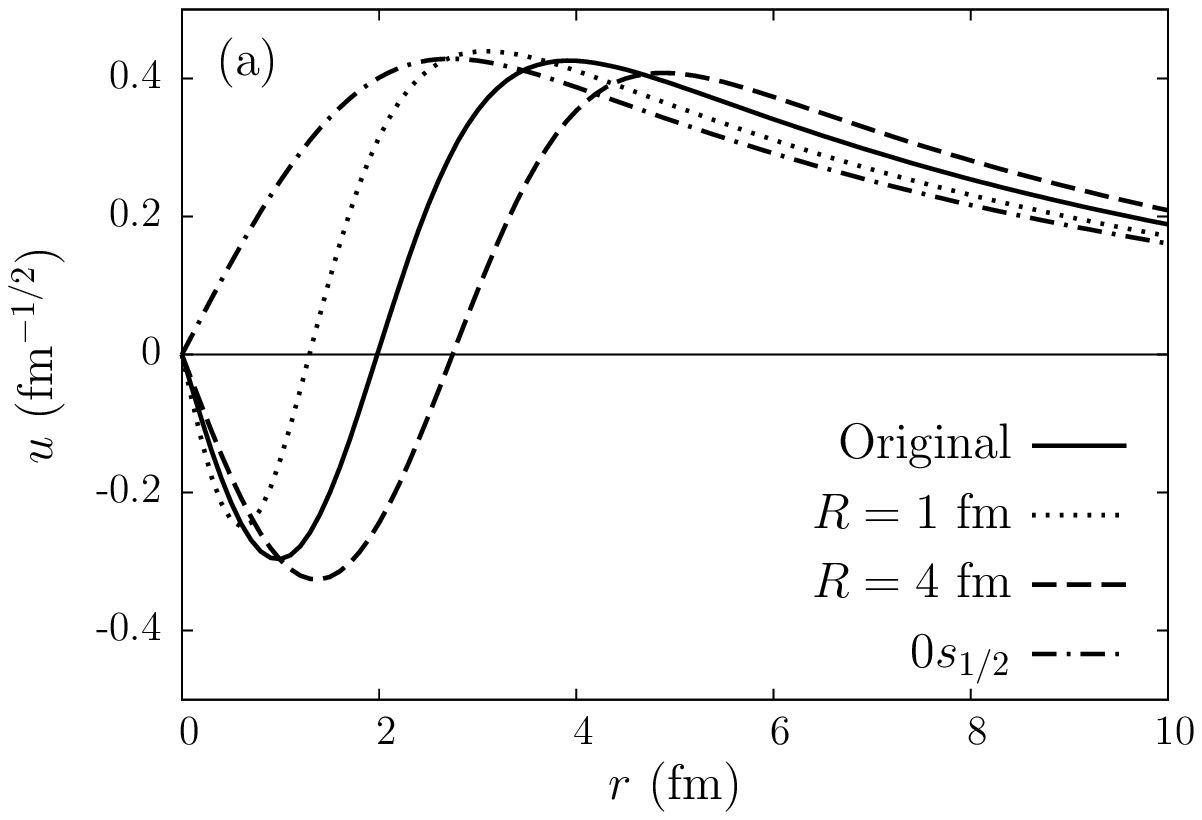}
\includegraphics[width=8.7cm]{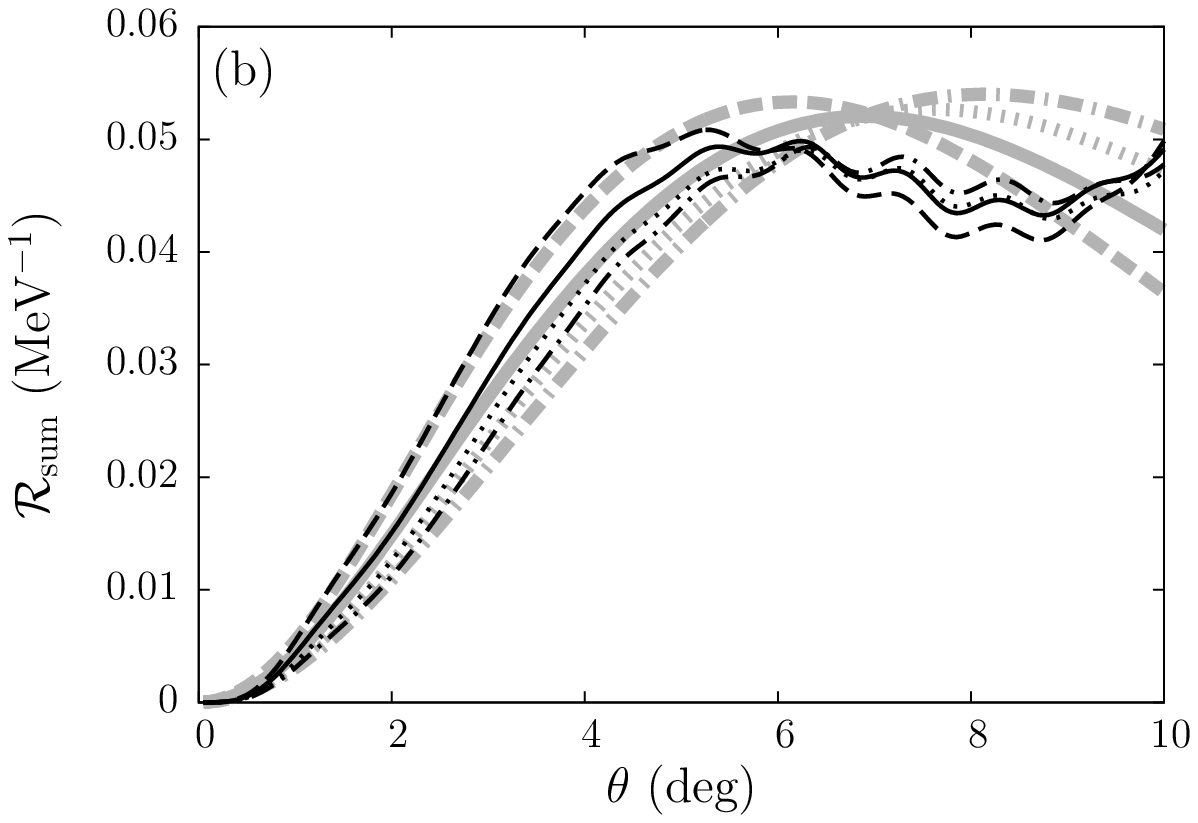}
\caption{Sensitivity of ratio ${\cal R}_{\rm sum}$ to the radial wave function of the projectile:
(a) radial wave functions of the initial $s_{1/2}$ states;
(b) corresponding DEA ratios ${\cal R}_{\rm sum}$ folded with experimental resolution (thin black lines) and their REB predictions (thick gray lines).
}\label{fig-wf}
\end{figure}

We now turn to the sensitivity of the ratio to details of the projectile radial wavefunction.
We consider various geometries for $V_{cn}$ in the $s$ wave and readjust the depth of the interaction to reproduce the physical neutron separation energy $E_0=-0.5$~MeV (see Appendix \ref{numerics}).
Namely we vary the radius and the number of nodes of the initial state.
The resulting radial wavefunctions are presented in  \Fig{fig-wf}(a).
We repeat DEA calculations for the reaction on Pb at 69 MeV/nucleon.
The corresponding ratios ${\cal R}_{\rm sum}$ folded with experimental resolution and their REB predictions are plotted in \Fig{fig-wf}(b).
At forward angles, i.e. in the range $1^\circ$ to $3^\circ$, the DEA ratios are in excellent agreement with their REB predictions.
In that region, the ratio is  proportional to the square of the asymptotic normalization coefficient (ANC).
At larger angles, the discrepancy between DEA and REB increases.
Nevertheless, the general behavior, and especially the ordering of the curves, is the same in both models.
In particular, the REB predicts a crossing of the curves, that actually takes place at $\theta\simeq 6^\circ$ in dynamical calculations.
At that angle, the ratio obtained with the initial $0s_{1/2}$ state overtakes the others although it corresponds to the smallest ANC.
This can only be explained if the ratio is sensitive to the internal part of the wave function at larger angles.
The ratio is thus able to probe different parts of the  radial wave function, depending on the scattering angle $\theta$.
It is one of the few reaction observables to display such a property.
Elastic-scattering and breakup cross sections are indeed purely peripheral, in the sense that they probe only the tail of the projectile wave function and not its interior \cite{CN06}.
Although these details do not affect the ratio as significantly as the binding energy and the angular momentum, we expect them to be observable in data with enough statistics.

\subsection{Choice of continuum energy $E$}\label{continuum}

\begin{figure}[t]
\center
\includegraphics[width=8.7cm]{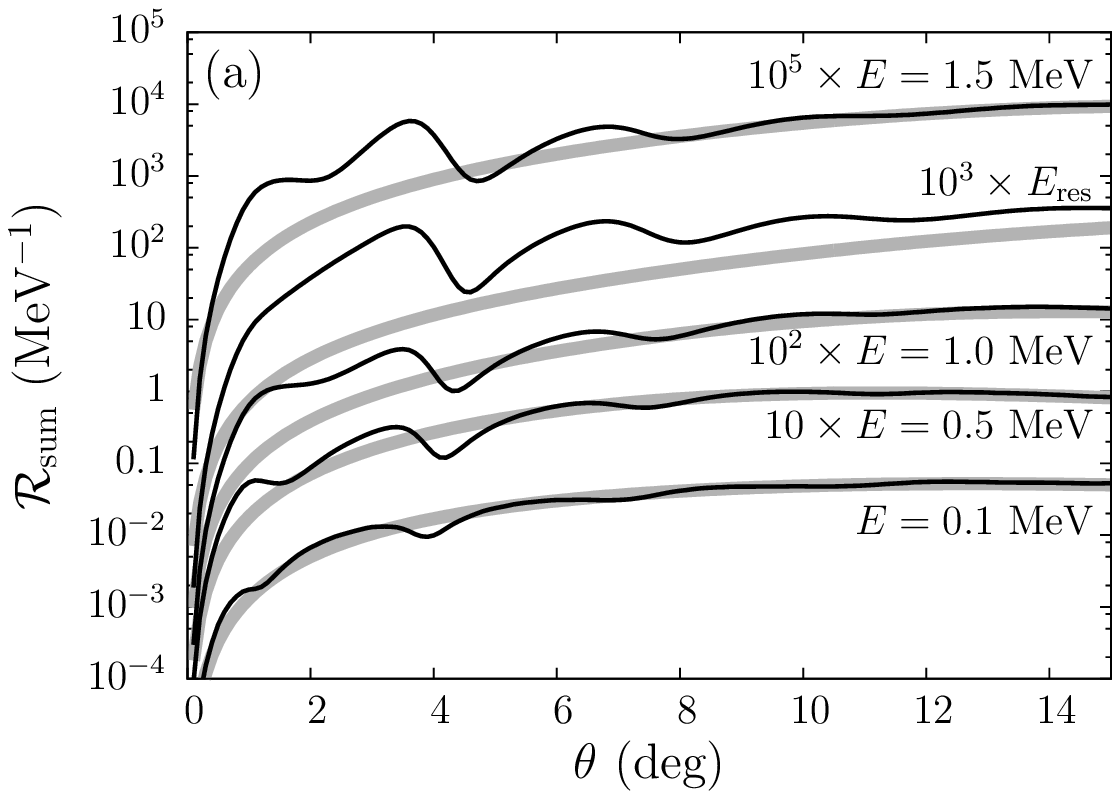}
\includegraphics[width=8.7cm]{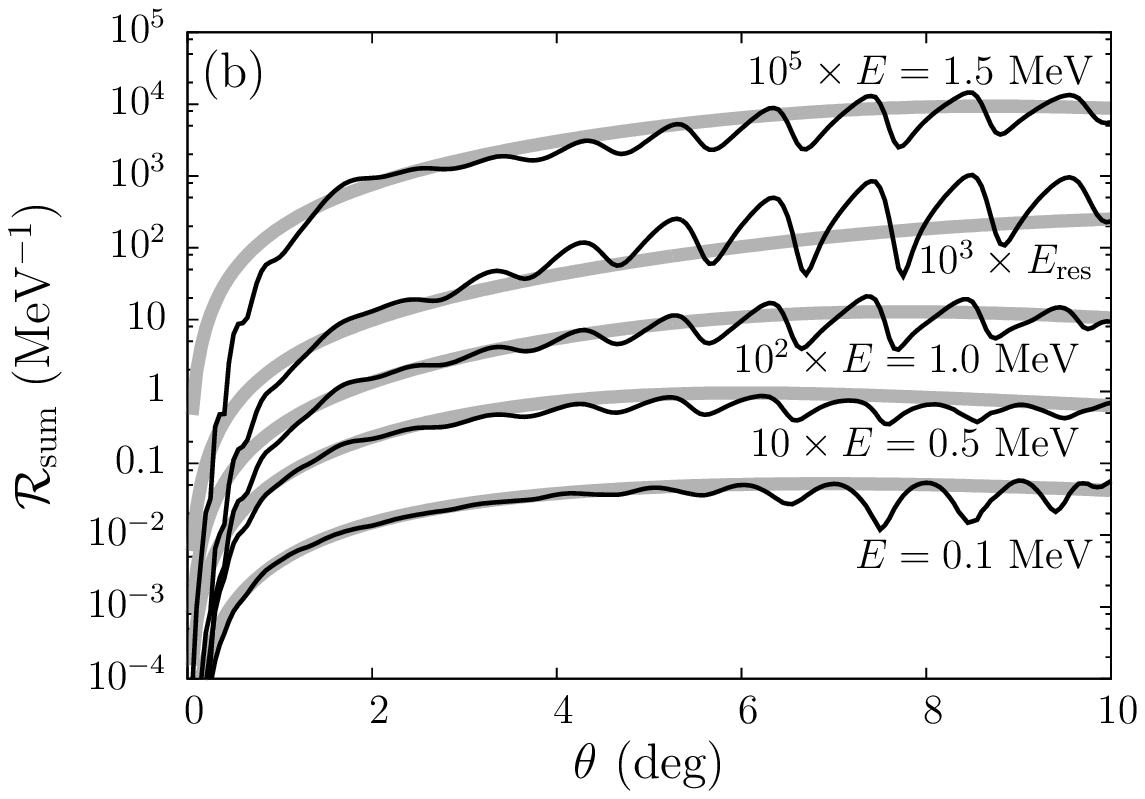}
\caption{Change of the ratio with the energy $E$ in the $c$-$f$ continuum.
(a) \ex{11}Be on C at 67~MeV/nucleon; (b) \ex{11}Be on Pb at 69~MeV/nucleon.
For convenience, each ratio has been multiplied by a factor.}
\label{fig-ebin}
\end{figure}

So far we have fixed the $^{10}$Be-$n$ relative energy in the final state to be $E=0.1$~MeV.
However ${\cal R}_{\rm sum}$ can be defined for any relative energy $E$ between the core and the halo neutron after breakup.
We now study the effect of this energy by considering $E=0.5$, 1.0, and 1.5~MeV.
In addition, we also consider the \ex{10}Be-$n$ energy $E_{\rm res}=1.274$~MeV, which corresponds to a $5/2^+$ resonance in the \ex{11}Be continuum.
In our model, that resonance is simulated by a $d5/2$ state \cite{capel04}.
The resulting ratios are shown in \Fig{fig-ebin} for a C target at 67~MeV/nucleon (a) and for a Pb target at 69~MeV/nucleon (b).
For readability, the ratios have been multiplied by powers of 10: 10 for $E=0.5$~MeV, $10^2$ for $E=1.0$~MeV, $10^3$ for $E_{\rm res}$, and $10^5$ for $E=1.5$~MeV.
As expected from the above analysis, the agreement between DEA calculations (solid lines) and REB predictions (thick gray lines) worsens with increasing continuum energy $E$.
On both targets, the residual oscillations increase at larger $E$.
This effect is caused by the shift due to $U_{nT}$, which increases with $E$ (see \Sec{vnt}).
The difference in the oscillatory pattern between the elastic scattering and the breakup angular distributions therefore increases at larger $E$, leading to more significant residual oscillations in the ratio.
This disagreement between DEA calculations and the REB predictions fully disappears when $U_{nT}$ is set to 0 (see Fig.~\ref{fig-vnt}).

For the heavy target, the rise of the DEA ratio at forward angles is slower compared to its REB prediction (see \Fig{fig-ebin}(b)).
As explained in \Sec{adiabatic}, this slower rise is due to the adiabatic approximation made in the REB.
Accordingly, the agreement between the dynamical calculation and the form factor $|F_{E,0}|^2$ at forward angles gets worse with increasing excitation energy.
This discrepancy is not observed on the light target for which the adiabatic approximation is well suited.

Of particular interest is the case where the final state is a resonance.
The calculations displayed in \Fig{fig-ebin} do not show unusual effects at $E_{\rm res}$.
The ratio rises slightly faster at large scattering angles and its residual oscillations are slightly larger than off-resonance.
These more significant departures from the REB prediction might be seen experimentally and could hence be used to spot resonant structures in the continuum of exotic nuclei.
However such a feature is more clearly observed in energy distributions measured after breakup on a light target \cite{fukuda04,capel04}.

This analysis shows that although the agreement between the DEA ratio and its REB prediction remains fair, the ratio method would be better applied at low energy $E$ in the projectile continuum and off resonance.
In the case studied here, the discrepancy remains small up to a \ex{10}Be-$n$ energy $E\sim0.5$~MeV.

\section{Conclusion and prospects}\label{conclusion}
A new observable for halo nuclei has been studied in detail.
It consists of the ratio of the breakup angular distribution and the summed angular distribution including elastic, inelastic and breakup.
We show that realistic calculations of this observable closely follow  predictions by the REB model. The latter neglects the neutron-target interaction and the excitation energy of the projectile.
In this work we have explored the small discrepancies that exist between dynamical calculations and their REB predictions:
the DEA ratio exhibits residual oscillations and rises more slowly than the REB at very forward angles for Coulomb-dominated processes.
The residual oscillations are caused by the additional kick the neutron feels due to its interaction with the target $U_{nT}$.
The difference observed at forward angles on heavy targets is due to the adiabatic approximation made in the REB, which is incompatible with the long range of the Coulomb interaction.
Nevertheless, these discrepancies remain small indicating that most of the dependence on the reaction mechanism is removed by taking this ratio of cross sections.
Therefore the cross section ratio is an optimal tool to study the structure of exotic nuclei.

We next analyze the structure information contained in the cross section ratio.
Our results show that the ratio is extremely sensitive to the binding energy of the halo and the angular momentum of the halo orbital.
While binding energy and angular momentum can be unequivocally extracted, an experimental error of a few percents would be necessary for learning about the details of the radial behavior of the halo.
Very accurate data could constrain simultaneously the asymptotic normalization coefficient as well as the internal behavior of the radial wave function.
One advantage of the ratio method is that it provides an additional control variable, the energy of the continuum bin, which can be tuned to be most appropriate to the case under study.
Although the method works best when the excitation in the continuum is small, we show that even at 1.5~MeV useful results can be obtained.

Thanks to its independence of the reaction mechanism, the cross section ratio  enables us to probe in-depth the structure of one-neutron halo nuclei and brings out information inaccessible to any other reaction observable.
According to our analysis, the ratio method works both on light and heavy targets.
It gives better results at high beam energy and for loosely-bound projectiles, for which the adiabatic approximation is more reliable.

\appendix

\section{Other ratios}\label{other}
We have also considered ratios obtained by integrating over the relative energy of the breakup fragments. This could be of interest due to low statistics of the energy distribution of the breakup fragments or other experimental constraints.
The resulting expressions are denoted by ${\cal R}_{\rm \int \!\!el}$ and ${\cal R}_{\rm \int \!\!sum}$ and are defined by
\beq
{\cal R}_{\rm \int \!\!el}(\ve Q) &=& \frac{\int d\sigma_{\rm bu}/dEd\Omega \; dE}{d\sigma_{\rm el}/d\Omega} \label{ratio-el2}\\
 &\stackrel{\rm (REB)}{=}&\frac{1-(|F_{0,0}(\ve Q)|^ 2+\sum_{i>0}|F_{i,0}(\ve Q)|^ 2)}{|F_{0,0}(\ve Q)|^2},
\eeqn{reb-el2}
and
\beq
{\cal R}_{\rm \int \!\!sum}(\ve Q) &=&\frac{\int d\sigma_{\rm bu}/dEd\Omega \;dE}{d\sigma_{\rm sum}/d\Omega}\label{ratio-sum2}\\
 &\stackrel{\rm (REB)}{=}&1-(|F_{0,0}(\ve Q)|^ 2+\sum_{i>0}|F_{i,0}(\ve Q)|^ 2).
\eeqn{reb-sum2}
In this case however, all configurations of the final breakup fragments are contributing.
As detailed in Secs.~\ref{adiabatic} and \ref{continuum}, the higher the relative energy between the breakup fragments $E$, the less valid is the adiabatic approximation assumed in the REB model. Therefore, we found that neither ${\cal R}_{\rm \int \!\!el}$ \eq{ratio-el2} nor ${\cal R}_{\rm \int \!\!sum}$ \eq{ratio-sum2} are useful. 

\section{Two-body interactions}\label{numerics}
Most of the calculations presented in this text have been performed using the DEA \cite{dea,dea2} with the description of \ex{11}Be developed in \Ref{capel04} and the optical potentials of \Ref{dea2}.
This appendix details the form-factors of these potentials and lists their parameters used in this study.

The projectile model is based on the two-body description presented in \Sec{model}.
The $V_{cn}$ potential contains a central term plus a spin-orbit coupling term
\beq
V_{cn}(\ve{r})=V_0 f(r,R_0,a)+V_{LS}\ \ve{l}\cdot\ve{s}\ \frac{1}{r}\frac{d}{dr}f(r,R_0,a),
\eeqn{eB1}
with the Woods-Saxon form factor
\beq
f(r,R_0,a)=\left[1+\exp\left(\frac{r-R_0}{a}\right)\right]^{-1}
\eeqn{eB2}
of radius $R_0$ and diffuseness $a$.
In \Eq{eB1}, $\ve{l}$ is the relative orbital angular momentum
between the core and the valence neutron, and $\ve{s}$ is the spin of the neutron.
For completeness, we list in \tbl{t1} the parameters of the $c$-$n$ potentials used in this analysis.
The first two rows contain the original potential of \Ref{capel04} for $^{11}$Be (for even and odd orbital angular momenta).
This potential reproduces the experimental binding energy $E_0=-0.5$~MeV in the $1s_{1/2}$ orbital to describe the $1/2^+$ ground state of \ex{11}Be.
For the $1/2^-$ excited state, we take the $0p_{1/2}$ bound state of the potential, which requires a different depth $V_0$ for even and odd partial waves.

The continuum wave functions appearing in \Eq{reb-ff2} are obtained with the $V_{cn}$ potential. Note that
the $5/2^+$ resonance at 1.274~MeV in the \ex{10}Be-$n$ continuum, is reproduced in the $d_{5/2}$ partial wave with the original potential of \tbl{t1} \cite{capel04}.
The other lines of \tbl{t1} contain the parameters used for the calculations presented in Figs.~\ref{fig-be}, \ref{fig-ang}, and \ref{fig-wf}.
The last line correspond to the potential used to describe $^{19}$C in the calculations presented in \Fig{fig-c19}.
The potentials are labeled as the corresponding curves in the figures.
In most of the $^{11}$Be calculations the parameters listed in \tbl{t1} are used only in the ground-state partial wave, the others,  including the continuum wave functions, being described by the ``Original'' potential.
Only in the $0p_{1/2}$ and $0d_{5/2}$ cases do we use the same potential in all partial waves.
In the $^{19}$C case, all partial waves are described using the same potential.

\begin{table}
\center
\begin{tabular}{ccccc}\hline\hline
 & $V_0$ & $V_{LS}$  &  $R_0$  & $a$\\
 & (MeV) &	(MeV fm\ex{2})&  (fm) &(fm)\\\hline
Original (even $l$) & 62.52 &    21.00    & 2.585 & 0.6\\
Original (odd $l$) & 39.74 &    21.00    & 2.585 & 0.6\\ \hline
$E_0=-50$~keV & 57.91 &    21.00    & 2.585 & 0.6\\
$E_0=-5$~MeV & 80.00 &    21.00    & 2.585 & 0.6\\
$R=1$~fm & 210.19 &    21.00    & 1 & 0.6\\
$R=4$~fm & 30.915 &    21.00    & 4 & 0.6\\
$0s_{1/2}$ & 11.191 &    21.00    & 2.585 & 0.6\\ \hline
$0p_{1/2}$ & 40.861 &    21.00    & 2.585 & 0.6\\
$0d_{5/2}$ & 74.162 &    0.00    & 2.585 & 0.6\\ \hline
$^{18}$C-$n$ & 42.161 & 14.00 & 3.24 & 0.62\\ \hline\hline
\end{tabular}
\caption{Parameters of the core-$n$ potentials \eq{eB1}.}\label{t1}
\end{table}

The nuclear part of the optical potentials used to simulate the interaction between the projectile constituents and the targets contains real and imaginary volume terms and an imaginary surface term
\beq
\lefteqn{U_{xT}=-V f(r,R_r,a_r)}\nonumber\\
&-&i \left[W f(r,R_i,a_i)+ W_d\ a_i \frac{d}{dr}f(r,R_i,a_i)\right].
\eeqn{e7}
The Coulomb part of $U_{cT}$ is simulated by the potential
due to a uniformly charged sphere of radius $R_C$.
The parameters of the optical potentials used in this study are listed in \tbl{t2}.
We follow \Ref{dea2} for the choices of most of these optical potentials.
For $U_{nT}$ we use the Becchetti and Greenlees parameterization \cite{BG69}, for the Pb target.
For the carbon target, we follow \Ref{capel04} and use the $p$-C potential of Comfort and Karp \cite{CK80}.
The \ex{10}Be-Pb potentials are obtained from $\alpha$-Pb potentials by merely rescaling the radius for an $A=10$ projectile \cite{Bon85,Gol73}.
For the \ex{10}Be-C potential, we follow \Ref{capel04} and use the potential developed by Al-Khalili, Tostevin and Brooke that fits the elastic scattering of \ex{10}Be on C at 49.3~MeV/nucleon \cite{ATB97}.
The \ex{18}C-Pb interaction is derived from the potential of \Ref{Bue84} that fits the \ex{13}C-Pb elastic scattering at 390~MeV.

\begin{table*}
\center
\begin{tabular}{ll|cccccccccc}\hline\hline
$P$ & $T$ & Energy & $V$ & $R_r$ & $a_r$ & $W$ & $W_d$ & $R_i$ & $a_i$ & $R_C$ & Ref.\\
      &  & (MeV/nucleon)&(MeV)&(fm)&(fm) &(MeV)& (MeV) &  (fm) & (fm)  & (fm)& \\\hline
$n$ & Pb &69 & 29.46 &6.93&0.75& 13.4 & 0     &7.47&0.58&-&\cite{BG69}\\
       &      &40 & 38.423 &6.93&0.75& 7.24 & 1.846&7.47&0.58&-&\cite{BG69}\\
       &      &100& 44.823 &6.93&0.75& 2.84 & 21.846&7.7&0.58&-&\cite{BG69}\\
       &      &67 & 29.78 &6.93&0.75& 13.18 & 0     &7.47&0.58&-&\cite{BG69}\\
       & C   &67 & 30.9 &2.75&0.623& 7.82 & 0     &3.18&0.667&-&\cite{CK80}\\
\ex{10}Be&Pb&69 & 79.5 & 8.08& 0.893 &36.5& 0 & 8.16& 0.846  & 8.08 &\cite{Bon85}\\
                 &   & 40 & 155 & 8.17 & 0.667 & 23.26&0& 9.42 & 0.733& 8.92 &\cite{Gol73}\\
                 &   &100& 71.5 & 7.92& 0.93  &57.6& 0 & 7.76& 0.934  & 8.08 &\cite{Bon85}\\
       & C   &67 & 123 &3.33&0.8& 65 & 0     &3.47&0.8&5.33&\cite{ATB97}\\ 
\ex{18}C&Pb&67 & 200.0 & 5.39& 0.9 &76.2& 0 & 6.58& 0.38  & 5.92 &\cite{Bue84}\\ \hline\hline
\end{tabular}
\caption{Parameters of the optical potentials \eq{e7} used to simulate the
interaction between the targets and projectile fragments $n$, and \ex{10}Be or \ex{18}C.
}\label{t2}
\end{table*}

\begin{acknowledgments}
We thank I.~J.~Thompson, B.~Tsang, N.~Timofeyuk, and the MoNA collaboration for interesting discussions on the subject.
This work was supported by the National Science Foundation grant PHY-0800026 and the Department of Energy under contract DE-FG52-08NA28552
and DE-SC0004087. R.~C.~J.\ is supported by the United Kingdom Science and Technology Facilities Council under 
Grant No. ST/F012012.
This text presents research results of the Belgian Research Initiative on eXotic nuclei (BriX),
program nr P7/12 on interuniversity attraction poles of the Belgian Federal Science Policy Office.
\end{acknowledgments}


\end{document}